\documentclass[onecollarge,natbib]{svjour2}
\bibpunct{[}{]}{;}{n}{}{,} 
\smartqed  
\usepackage{graphicx}
\usepackage{amsmath}
\usepackage{subfigure}
\usepackage{xcolor}
\newcommand{\leftsub}[2]{\vphantom{#2}_{#1}{#2}}

\journalname{Few-Body Systems (EFB22)}
\begin{document}

\title{Progress on Light-Ion Fusion Reactions with Three-Nucleon Forces \thanks{Prepared in part by LLNL under Contract DE-AC52-07NA27344. We acknowledge support from the U. S. DOE/SC/NP (Work Proposal No. SCW1158),  from the Deutsche Forschungsgemeinschaft through contract SFB 634, from the Helmholtz International Center for FAIR within the framework of the LOEWE program launched by the State of Hesse, from the BMBF through contract 06DA7074I and from the NSERC Grant No. 401945-2011. TRIUMF receives funding via a contribution through the Canadian National Research Council. Computing support for this work came from the LLNL institutional Computing Grand Challenge program, the J\"ulich Supercomputing Center, the LOEWE-CSC Frankfurt, and the National Energy Research Scientific Computing Center supported by the Office of Science of the U.S. Department of Energy under Contract No. DE-AC02-05CHH11231.}
}


\author{{G. Hupin} \and {S. Quaglioni}\and {J. Langhammer} \and {P. Navr\'atil} \and{A. Calci}\and {R. Roth}
}


\institute{G. Hupin \at Lawrence Livermore National Laboratory, P. O. Box 808, L-414, Livermore, California 94551, USA \\
              Tel.: +1-925-422-1608\\
              Fax: +1-925-422-5940\\
              \email{hupin1@llnl.gov}           
           \and
           S. Quaglioni \at Lawrence Livermore National Laboratory, P. O. Box 808, L-414, Livermore, California 94551, USA \\
           \and
           J. Langhammer \at Institut f\"ur Kernphysik, Technische Universit\"at Darmstadt, D-64289 Darmstadt, Germany \\
           \and
           P. Navr\'atil \at TRIUMF, 4004 Wesbrook Mall, Vancouver, British Columbia, V6T 2A3, Canada \\
           \and
           A. Calci \at Institut f\"ur Kernphysik, Technische Universit\"at Darmstadt, D-64289 Darmstadt, Germany \\
           \and
           R. Roth \at Institut f\"ur Kernphysik, Technische Universit\"at Darmstadt, D-64289 Darmstadt, Germany \\
}

\date{Received: date / Accepted: date}

\maketitle

\begin{abstract}
The description of structural and dynamical properties of nuclei starting from the fundamental interaction between nucleons has been a long-standing goal in nuclear physics. The {\em ab initio} No-Core Shell Model combined with the Resonating-Group Method (NCSM/RGM) is capable of addressing both structural and reaction properties of light-nuclei. While promising results have already been achieved starting from a two-body Hamiltonian, a truly realistic prediction of nuclear observables requires the treatment of the three-nucleon interaction. Using similarity-renormalization-group evolved two- and three-nucleon interactions, we will present recent applications to $n$-$^4$He scattering process when accounting for the chiral two- plus three-nucleon interaction versus the chiral two-nucleon interaction. We compare our results to phase shifts obtained from $R$-matrix analysis of data up to $16$ MeV neutron energy, below the $d$-$^3$H threshold.
\end{abstract}

\section{Introduction}
\label{intro}

Nuclei are self-bound systems consisting of composite objects, protons and neutrons. The interaction between nucleons is complex as it arises from the underlying theory of quantum chromodynamics. Nevertheless, understanding this interaction is essential to grasp the static and dynamical properties of nuclei. These properties, in turn, constitute an important piece of the puzzle that created and drives the evolution of our universe such as low-energy light-nuclei fusion reactions occurring in stars. We can study the nuclear interaction within the framework of the No-Core Shell Model combined with the Resonating Group Method (NCSM/RGM). This method is capable of addressing both static and dynamical properties of nuclei, until recently using only two-nucleon ($NN$) interactions. Recently, we have studied the inclusion of a three-nucleon ($3N$) force, in particular the chiral N$^2$LO of Ref.~\cite{Navratil2007}, in the formalism and its effects on the phase shifts of the $n$-$^4$He elastic scattering~\cite{Hupin2013}; we present here a selection of those results.

\section{Formalism}
\label{Formalism}

The many-body wave function of the $A$-nucleon system,
\begin{equation}
|\Psi^{J^\pi T}\rangle = \sum_{\nu \nu'} \int dr dr^\prime ~ r^{2} r^{\prime\, 2}  ~  [{\cal N}^{-\frac12}]^{J^\pi T}_{\nu\nu^\prime}(r,r^\prime)\frac{\chi^{J^\pi T}_{\nu^\prime}(r^\prime)}{r^\prime} ~ {\mathcal A}_{\nu}|\Phi^{J^\pi T}_{\nu r}\rangle \, , \label{eq:ansatz}
\end{equation}
is expanded over a continuous set of cluster basis states, $| \Phi^{J^\pi T}_{\nu r} \rangle$. The cluster basis consists of translationally-invariant states describing two nuclei, a target of mass number $(A{-}a)$ and an $a$-nucleon projectile, traveling in a $^{2s+1}\ell_J$ partial wave of relative motion (with $s$ the channel spin, $\ell$ the relative orbital angular momentum, and $J$ the total angular momentum of the system). The operator ${\mathcal A}_{\nu}$ enforces the antisymmetrization of nucleons that pertain to different clusters. The coefficients $\chi^{J^\pi T}_{\nu'}(r')$ represent continuous linear variational amplitudes. These are determined by solving the orthogonalized NCSM/RGM equations (we refer the interested reader to Sec.\ II.E of Ref.~\cite{Quaglioni2009}):
\begin{equation}
\sum_{\gamma\gamma'\nu'} \! \int \!  dydy'dr^\prime y^2 y'^2 r^{\prime\,2} [{\mathcal N}^{-\frac12}]^{J^\pi T}_{\nu\gamma}(r,y)~{\mathcal H}^{J^\pi T}_{\gamma\gamma'}(y,y')~[\,{\mathcal N}^{-\frac12}]^{J^\pi T}_{\gamma'\nu'\,}(y',r') \frac{\chi^{J^\pi T}_{\nu^\prime} (r^\prime)}{r^\prime}= E\,\frac{\chi^{J^\pi T}_{\nu} (r)}{r}  \, , \label{eq:RGMeq}
\end{equation}
where $E$ is the total energy in the center of mass (c.m.) frame. The norm and Hamiltonian integration kernels, $ {\mathcal N}^{J^\pi T}_{\nu\nu^\prime\,}(r,r^\prime)$ and ${\mathcal H}^{J^\pi T}_{\nu\nu^\prime\,}(r,r^\prime)$ respectively, are the  overlap and matrix elements of the Hamiltonian with respect to the cluster basis, i.e.
\begin{equation}
{\mathcal N}^{J^\pi T}_{\nu^\prime \nu}(r^\prime, r) =  \langle \Phi^{J^\pi T}_{\nu^\prime r^\prime} | {\mathcal A}_{\nu^\prime} {\mathcal A}_{\nu} | \Phi^{J^\pi T}_{\nu r} \rangle\, ~ \textrm{and}~
{\mathcal H}^{J^\pi T}_{\nu^\prime \nu}(r^\prime, r) =  \langle\Phi^{J^\pi T}_{\nu^\prime r^\prime} |  {\mathcal A}_{\nu^\prime}  H  {\mathcal A}_{\nu} | \Phi^{J^\pi T}_{\nu r} \rangle \, .
\label{eq:NH-kernel}
\end{equation}
The microscopic $A$-nucleon Hamiltonian, $H$, consists of the relative kinetic energy, the intrinsic Hamiltonians of both clusters and the inter-cluster nuclear plus Coulomb interaction. The norm and Hamiltonian kernels are obtained from the eigenvectors of the target and projectile after expanding on a Harmonic Oscillator (HO) NCSM basis controlled by the parameter $N_{\rm max}$. In this work, the nuclear interaction accounts for the two- and three-nucleon interaction, $V^{NN}$ and $V^{3N}$, respectively. For instance, the three-nucleon force contributions to the Hamiltonian kernel in the model-space are
\begin{eqnarray}
{{\mathcal V}^{3N}_{\nu^\prime n'\nu n}}&=& \phantom{-} \frac{(A{-}1)(A{-}2)}{2} ~~ \leftsub{\rm SD}{\langle}\Phi^{J^\pi T}_{\nu^\prime n^\prime} |  {V}^{3N}_{A-2,A-1,A}(1-2  P_{A-1,A}) | \Phi^{J^\pi T}_{\nu n}\rangle_{\rm SD} \label{pot-NNN-direct} \\
 & &- \frac{(A{-}1)(A{-}2)(A{-}3)}{2} ~~ \leftsub{\rm SD}{\langle}\Phi^{J^\pi T}_{\nu^\prime n^\prime} |   P_{A-1,A} {V}^{3N}_{A-3,A-2,A-1} |\Phi^{J^\pi T}_{\nu n} \rangle_{\rm SD}\, ,\label{pot-NNN-ex}
\end{eqnarray}
in the Slater-determinant (SD) basis of Eq~(31) of Ref.~\cite{Quaglioni2009} where $n'$ and $n$ are the radial quantum number of the HO expansion. As in the two-nucleon case of Ref.~\cite{Quaglioni2009}, we identify a direct [Eq.~\ref{pot-NNN-direct}] and an exchange [Eq.~\ref{pot-NNN-ex}] term.

\section{Applications}
\label{applications}

Applications of the NCSM/RGM approach for the description of single-nucleon projectile collisions have already led to very promising results based on $NN$ interactions~\cite{Quaglioni2008,Quaglioni2009}. The nucleon-$^4$He is an ideal testing ground to investigate the importance of the chiral $3N$ force on low-energy reaction observables. It consists in a single open channel up to fairly high energy and, at the same time, it is sensitive to the strength of the ($NN$+$3N$) spin-orbit force as demonstrated in earlier studies of the $^2$P$_{3/2}$ and $^2$P$_{1/2}$ scattering phase-shifts~\cite{Nollett2007,Quaglioni2008,Quaglioni2009}. For this study~\cite{Hupin2013}, we work with the chiral N$^3$LO $NN$ interaction of Ref.~\cite{Entem2003} and chiral N$^2$LO $3N$ interaction of Ref.~\cite{Navratil2007} both evolved with the Similarity-Renormalization Group (SRG) method~\cite{Hergert2007,Bogner2007} that softens the short-range repulsion of the nuclear interaction. It should be noticed that, already starting from an initial $NN$ Hamiltonian, the SRG procedure generates induced $3N$ forces that have to be taken into account. We denote the two-body portion of the SRG-evolved chiral $NN$ interaction as $NN$-only and label $NN$+$3N$-full results in which induced and SRG-evolved chiral $3N$ forces are also included. A study of the dependence of the phase-shifts with respect to the model-space parameters shows overall good convergence properties for $N_{\rm max}{=}13$ HO shells and including the first six excited states of the $^4$He, to account for the polarization of the target~\cite{Hupin2013}. 

\begin{figure}
\begin{minipage}[c]{.48\linewidth}
\centering
\includegraphics[width=.99\linewidth]{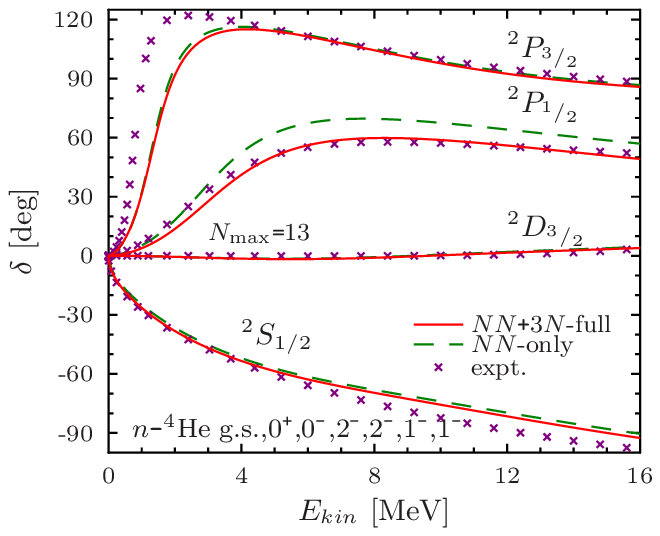}
\caption{(Color online) Comparison between our calculated $NN$-only (dashed green lines) and $NN$+$3N$-full (solid red lines) $n$-$^4$He phase-shifts ($^2$S$_{1/2}$, $^2$P$_{1/2}$, $^2$P$_{3/2}$ and $^2$D$_{3/2}$ partial waves) and phase shifts obtained from an accurate $R$-matrix analysis (purple crosses)~\cite{Hale}. We account for the polarization of the target by including the first six low-lying states of the $^4$He. Additional parameters are $N_{\rm max}{=}13$, $\lambda{=}2.0$ fm$^{-1}$ and $\hbar \Omega{=}20$ MeV.}
\label{fig:1}
\end{minipage}
\hfill
\begin{minipage}[c]{.48\linewidth}
\centering
 \includegraphics[width=.99\linewidth]{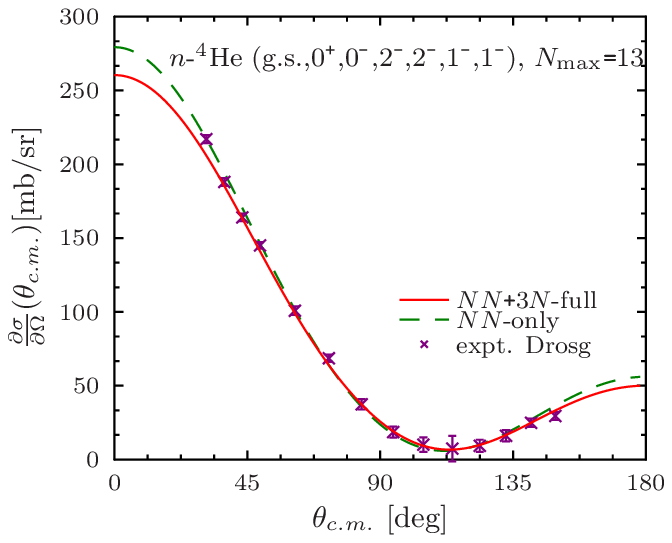}
\caption{(Color online) Comparison between our calculated $NN$-only (dashed green lines) and $NN$+$3N$-full (solid red lines) differential cross-section of the $n$-$^4$He collision and data (purple crosses) of Ref.~\cite{Drosg1978} at incident neutron energy of $17.6$ MeV. Additional parameters are identical to the one of Fig. \ref{fig:1}.}
\label{fig:2}
\vspace{12mm}
\end{minipage}
\end{figure}

In figure \ref{fig:1}, we compare our low-energy $n$-$^4$He scattering phase shifts to phase shifts obtained from an accurate $R$-matrix analysis of $^5$He data (crosses)~\cite{Hale}. The model-space parameters are $N_{\rm max}{=}13$, the SRG flow parameter $\lambda{=}2.0$ fm$^{-1}$, the HO frequency $\hbar \Omega{=}20$ MeV and the first six low-lying states of $^4$He are included. The phase shifts obtained with the $NN$+$3N$-full Hamiltonian are shown as solid red lines while those obtained from the $NN$-only Hamiltonian are given as dashed green lines. When the induced and SRG-evolved chiral $3N$ forces are included, we find a fairly good reproduction of the experimental phase-shifts for all the partial waves at incident neutron energy below the $d$-$^3$H threshold. 
In particular, the induced $3N$ interaction is responsible for a significant overall reduction of the P-wave phase shifts compared to the $NN$-only case (not shown in the figure). However, the initial $3N$ interaction increases the spin-orbit splitting between the P waves, bringing the theoretical results closer to experiment.
For energies around the resonance position, both $NN$-only and $NN$+$3N$-full $^2P_{3/2}$ partial waves miss the experimental resonance energy at $0.78$ MeV. The reason for this disagreement could be twofold: First, the need for a more complex spin-orbit structure of the $3N$ force cannot be ruled out \cite{Nollett2007}. Second, the NCSM/RGM model space may still be insufficient for grasping $A$-body short-range correlations in the $^2P_{3/2}$ channel. To describe those correlations, we have included the first six excited states of $^4$He corresponding to $24$ MeV of excitation energy, however accounting for, e.g., the coupling to the $d$-$^3$H channel that opens experimentally at $17.63$MeV would lead to a more complete description of the scattering process. A way to overcome this difficulty is to treat on the same footing the long-range cluster correlations and short-range $A$-body correlations, such as in the No-Core Shell Model with Continuum (NCSMC) of Refs.~\cite{Baroni2013,Baroni2013a}. Despite of this, as illustrated in Fig.~\ref{fig:2}, we observed a fairly good reproduction of the $n$-$^4$He experimental differential cross-section, in general, using either $NN$-only or $NN$+$3N$-full Hamiltonians, while larger discrepancies with data remain for the $A_y$ polarization observable of this system, even if the chiral $3N$ interaction improves the agreement.

\section{Conclusion}
\label{conclusion}

We have presented an outline of the NCSM/RGM, an {\em ab initio} many-body approach capable of describing static and dynamical properties of nuclei. In the binary cluster formalism, the eigenstates of the target and projectile are combined into a channel basis expansion and complemented with realistic two- and three-nucleon interactions. We discussed the inclusion of the three-nucleon force in the NCSM/RGM formalism using the simple but instructive case of the elastic $n$-$^4$He scattering. The $3N$-force contribution plays an important role in reproducing the experimental phase-shifts, in particular the spin-orbit splitting between the $P$ waves. This work is the first step towards high-precision nuclear reaction calculations with $NN$ and $3N$ interactions~\cite{Hupin2013}.

%
%


\bibliographystyle{spbasic}
\bibliography{biblio.bib}   

%
%

\end{document}